# A Semantic Without Syntax

Farzad Didehvar[1]
didehvar@aut.ac.ir

Department of Mathematics &Computer Science,
Amir Kabir University

**Abstract:** *Here, by introducing a version of "Unexpected hanging paradox" we try to open a new way and a new explanation for paradoxes, similar to liar paradox. Also, we will show that we have a semantic situation which no syntactical logical system could support that. In the end, we propose a claim as a question. Based on this claim, having an axiomatic system for computability theory is not possible. In fact we will show that the method applied here could yields us as a generalized result, some Theories like Physic is not axiomatizable.*

**Introduction:** In [4] by applying a version of "Unexpected hanging paradox" we showed that there is a proof which doesn't show the truth. Here, we try to formalize this problem. First, we explain the scenario of paradox again.

**Scenario:**

In [4] we represented the following version of "Unexpected hanging paradox" :

Once upon a time a logician was arrested. He was judged in a court. As all knows the interest of our logician was to arguing, he was sentenced as following by judges:

"You will be executed in the next week if and only if you don't conclude logically in a written form at the time of execution orthe days before the execution that you will be executed".

The logician started the arguing and he wrote the result of his arguing as following: (Later on he sent this message to his lawyer)

I prove that I will not be executed the next week. Suppose that it is so:

1) I will not be executed on Friday, if so since till Friday I will not be executed; I conclude that I will be executed in Friday.
2) Analogously , I will not be executed on Thursday, If so since by (1) I will not be executed on Friday
   And till Wednesday I am not executed so I conclude that I will be executed on Wednesday. But as judge says I can't conclude that (logically and in a written form). So I will not be executed at that time. So I will not be executed
3) Analogously, I will not be executed on Wednesday, if so by (1) and (2) I will not be executed on Friday and on Thursday ….

7) I will not be executed on Saturday. By 1, 2,…6 and I conclude that I will be executed on Saturday

By what judge says I will not be executed.*

---

1. 424 Hafez Ave, Tehran, Iran, 15875-4413. +98 (21) 64540

I stop to conclude more and thinking about this subject more.

He sent this message to his lawyer, and in each day he sent a short message based on that he confirmed his conclusion. More exactly he said: I will not be executed tomorrow. On Wednesday the lawyer had no more messages.

As he understood, the logician was executed on Tuesday! With an open surprised eyes at the time of execution. He claimed injustice, and he exhibited the message of poor logician to journals and the court.

The court said:

The logician proved that he will not be executed, a true proof. We executed him on Tuesday, and as he claimed he didn't conclude in a written form that he will be executed on Tuesday, on the contrary he claimed that "he will not be executed in that day". In fact and in other words, he proved in his message that by accepting what the judge said as a true claim, he would not be executed in that week. More formally,

P: If we consider what the judge said as a true claim.

q: He would not be executed in that week.

His proof (p|---q)  is true as a proof, but it doesn't show the Truth.

In this paper we defend the above claim and we try to show   how we could develop this idea.

We should note, we don't claim that the above result is not proposes to be a solution for unexpected hanging paradox in all its versions. It is simply considered as the only way to explain this version of "Unexpected hanging paradox". Later on, we will know the above result as a possible explanation for the other versions of this paradox and some other paradoxes.

To formalize the above proof, it is sufficient to consider that A (prisoner) as a Turing Machine which could utter the code of a phrase, like [$\varphi$]. We have given formalization for the above paradox, and we note that any formalization should conclude this formalization and all suffer that they are contradictory.

**An Explanation and Conclusion:**

In the above system (formalization of paradox) there is a contradiction, in brief the judges claim that the prisoner will be executed next week, but the prisoner prove that they will not. So this system is a contradictory system, but at the same time we have a semantic for this system. So we have a contradictory system which has a semantic.
 In other word, we have semantics which no consistent syntactical system supports them.
As a result, our proof in above does not show any truth. So there are some intrinsically deficiencies in modeling and formalizing the proofs. In other word, formal systems are not able to support such semantic situations. Clearly, this opens a way to a new explanation for some paradoxes similar to liar paradox, as following:
In such paradoxes the proofs don't show the truth. Since there is no consistent syntactical system to support the related semantic.

This would be considered as the central result and idea of this paper. As the last word, we propose the following question, in the subject of Computability theory.

*Question:* Is the following a true claim?
In the above formalism, we could consider A as a Turing machine. By a slightly modification in the formalism
We are able to replace A by [A] (code of A), in a similar way $\varphi$ by $[\varphi]$. So the above problem has a formalism in the scope of Computability Theory. If one of the real goal of Computability Theory is to explain the real situations,
It should explain the above semantic situation. But by above explanation we know that no axiomatization is able to afford this, since any axiomatization falls in contradictory. So Computability Theory is not axiomatizable.

Forthcoming, it is notable to see three points here:

1. As formalizing of the problem shows we are able to rearticulate the paradox such that the concept of time would be eliminated.
2. In addition to Computability Theory the above claim is true for some other Theories as Physics and Mathematics as a whole.
3. One of the most important facts here is: In this paradox there are no debatable and fully controversial objects and concepts like *infinity*.

There is a weaker but less controversial result here. Any axiomatization of Computability Theory implies the inability of Computability Theory to explain and to describe the above paradox. But our Mind, by grasping semantic and syntax around this problem is able to understand the gist of the situation around this paradox.
So either Computability Theory is not axiomatizable or our Mind is not equivalent to a Turing machine.

Finally, It is notable to say, in all above we could replace Turing machine by any machine stronger than Turing machine.
It is certain that, the above issues introduce axiomatization approach as a weak approach to study in so many subjects.

*Acknowledgement:*
I would like to thank the staff of Amir kabir University, and department of Mathematics and Computer science.
I should thank M.Soltani, S.Soltani and A.Didehvar who helped me in some efforts around this paper. Also, I should Thank people who I thanked in [4].